\documentclass{mn2e}
\input{epsf}

\title[The anatomy of Leo I]
{The anatomy of Leo I: how tidal tails affect the kinematics}

\author[E. L. {\L}okas et al.]
    {Ewa L. {\L}okas,$^{1}$ Jaros{\l}aw Klimentowski,$^{1}$ Stelios Kazantzidis$^{2}$ and
	Lucio Mayer$^{3,4}$
    \\
    \\
    $^1$Nicolaus Copernicus Astronomical Center, Bartycka 18,
    00-716 Warsaw, Poland\\
    $^2$Center for Cosmology and Astro-Particle Physics;
        and Department of Physics; and Department of Astronomy, \\
    The Ohio State University, Physics Research Building, 191 West Woodruff Avenue, Columbus, OH 43210, USA\\
    $^3$Institute for Theoretical Physics, University of Z\"urich, CH-8057 Z\"urich, Switzerland\\
    $^4$Institute of Astronomy, Department of Physics, ETH Z\"urich, Wolfgang-Pauli
    Strasse, CH-8093 Z\"urich, Switzerland }
\begin{document}

\maketitle

\begin{abstract}
We model the recently published kinematic data set for Leo I dSph galaxy
by fitting the solutions of the Jeans equations to the
velocity dispersion and kurtosis profiles measured from the data. We demonstrate that
when the sample is cleaned of interlopers the data are consistent with the assumption that
mass follows light and isotropic stellar orbits with no need for an extended dark matter halo.
Our interloper removal scheme does not
clean the data of contamination completely, as demonstrated by the rotation curve of Leo I.
When moving away from the centre of the dwarf, the rotation appears to be reversed.
We interpret this behaviour using the results of an $N$-body simulation of a dwarf galaxy
possessing some intrinsic rotation, orbiting
in the Milky Way potential and show that it can be reproduced if the galaxy is viewed almost along the
tidal tails so that the leading (background) tail contaminates the western part of Leo I
while the trailing (foreground) tail the eastern one. We show that this configuration
leads to a symmetric and Gaussian distribution of line-of-sight velocities.
The simulation is also applied to test our modelling method on mock data sets. We demonstrate
that when the data are cleaned of interlopers and the fourth velocity moment is used the
true parameters of the dwarf are typically within $1\sigma$ errors of
the best-fitting parameters. Restricting the
fitting to the inner part of Leo I our best estimate for the anisotropy is
$\beta=-0.2^{+0.3}_{-0.4}$ and the total mass $M=(4.5 \pm 0.7) \times 10^7$ M$_\odot$.
The mass-to-light ratio including the errors in mass, brightness and distance
is $M/L_V=8.2 \pm 4.5$ solar units.
\end{abstract}


\section{Introduction}

The Leo I dwarf spheroidal (dSph) galaxy discovered by Harrington \& Wilson (1950) is one of
the brightest and most distant members of the Local Group dSph galaxy population. Its large
distance and significant velocity directed away from the Milky Way make its dynamical status still
unclear, both in terms of whether it is bound to either the Milky Way or M31 (Byrd et al. 1994)
and to what extent its internal dynamics may be affected by tidal interactions with the host galaxy.

The heliocentric velocity of Leo I was first determined from a single carbon star by
Aaronson, Hodge \& Olszewski (1983) which was later followed by measurements for red giants
by Suntzeff et al. (1986). Mateo et al. (1998) estimated with good accuracy
radial velocities for 33 red giants which allowed them to determine for the first time the
galaxy's velocity dispersion and mass-to-light ratio ($M/L$) of 3.5--5.6 solar units in $V$-band. Although
lower than in other dSph galaxies, this value of $M/L$ indicates the presence of a significant
amount of dark matter given the relatively young stellar population of Leo I (Lee et al. 1993;
Caputo et al. 1999; Gallart et al 1999).

\begin{figure}
\begin{center}
    \leavevmode
    \epsfxsize=8.2cm
    \epsfbox[110 50 470 410]{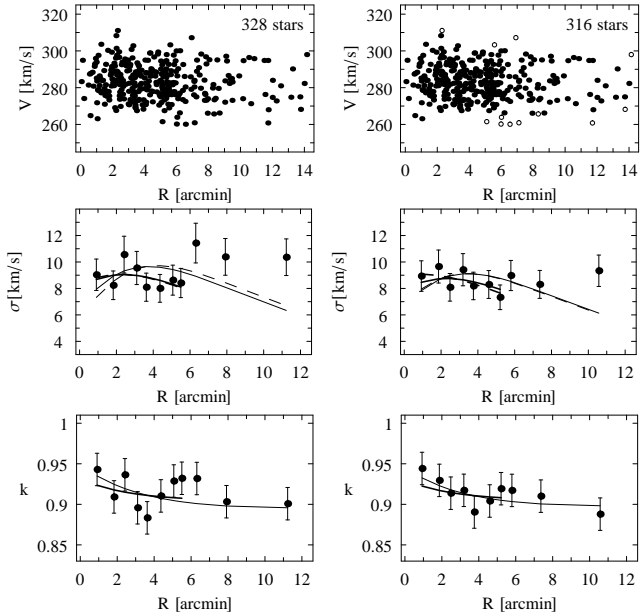}
\end{center}
\caption{Upper panels: the kinematic samples of Leo I stars used in the modelling. The left panel shows
the velocities versus projected distance from the galaxy centre $R$ for
the original sample of 328 stars identified as members by M08, while the right one presents the same for
the sample of 316
stars obtained by rejection of interlopers (open circles). The middle (lower) panels show the velocity
dispersion (kurtosis) profiles obtained from the corresponding samples with binning $9 \times 30 + 2 \times 29$
for the sample of 328 stars and $4 \times 31 + 6 \times 32$ for the sample of 316 stars. The dashed lines
show the best-fitting dispersion profiles when only the dispersion is fitted, while the solid lines plot the
best-fitting profiles of the moments when the dispersion and kurtosis are fitted simultaneously.
The thinner lines were obtained from fitting all data points, the thicker ones with the three outer data points
rejected.}
\label{moments}
\end{figure}

During the past year, three studies on Leo I dynamics have appeared in the literature:
Koch et al. (2007, hereafter K07), Sohn et al.
(2007, hereafter S07) and Mateo, Olszewski \& Walker (2008, hereafter M08). Each presented new
kinematic measurements for Leo I stars and discussed their interpretation. No consistent image of
the galaxy dynamics however emerged from these studies and their conclusions were on many points
contradictory. First, while S07 and M08 estimated the mass-to-light ratio of 10 solar units in the V-band
or lower, K07 found a value as high as 24. In addition, M08 and K07 claimed that the kinematic data are
inconsistent with a simple hypothesis that mass follows light and require an extended dark matter halo.
Second, all three investigations reported the detection of
rotation at some level, but different interpretations of this finding were given by S07 and M08:
M08 concluded that the western part of Leo I showing stellar velocities positive
with respect to the mean is affected by the leading tidal tail and the eastern part with velocities
below the mean by the trailing tail; the interpretation of S07 placed the corresponding tails in
opposite directions. Third, while S07 found the velocity distribution of their kinematic sample
to be asymmetric and interpreted it as a signature of tidal interaction, the distribution of the stellar sample
of M08 is symmetric and Gaussian-like.
The purpose of this work is to explain the differences and propose a detailed model for the
origin of the kinematic properties of Leo I.

The paper is organized as follows. In section 2 we present detailed models of Leo I kinematics
assuming that mass follows light and using the data set of M08. The data are modelled by fitting
the velocity dispersion and kurtosis profiles. We also discuss the possible contamination
of the data by stars from the tidal tails. We demonstrate that this contamination manifests itself
not only in the overestimated values of the outer dispersion data points but also in the shape of the
galaxy's rotation curve. In section 3 we use a collisionless $N$-body simulation of a dwarf galaxy orbiting
in the Milky Way potential to propose a detailed model of Leo I including the orientation of its tidal tails
with respect to the observer and show that such a configuration leads to a symmetric velocity
distribution. We also use the simulation to generate mock data sets and model them in order the verify
the reliability of our method. The discussion, including a detailed comparison with earlier work,
follows in section 4.

\section{Modelling of Leo I}

\begin{table}
\caption{Adopted parameters of Leo I. }
\label{adopted}
\begin{center}
\begin{tabular}{ll}
parameter & value \\
\hline
centre                                   & RA=$10^{\rm h} 08^{\rm m} 27^{\rm s}$ \\
					 & Dec=$+12^\circ18'30''$  \\
distance modulus $(m-M)_0$               & $22.02 \pm 0.13 $    \\
distance $D$                             & $254 \pm 15$ kpc \\
apparent magnitude $m_V$                 & $10.0 \pm 0.3$  \\
absolute magnitude $M_V$                 & $-12.02 \pm 0.43$  \\
luminosity $L_V$                         & $(5.5 \pm 2.2) \times 10^6 L_{\sun}$ \\
S\'ersic radius $R_{\rm S}$              & 5.0 arcmin  \\
S\'ersic parameter  $m$                  & 0.6 \\
major axis PA                            & $79^\circ$  \\
\hline
\end{tabular}
\end{center}
\end{table}

\begin{figure}
\begin{center}
    \leavevmode
    \epsfxsize=8cm
    \epsfbox[100 25 440 710]{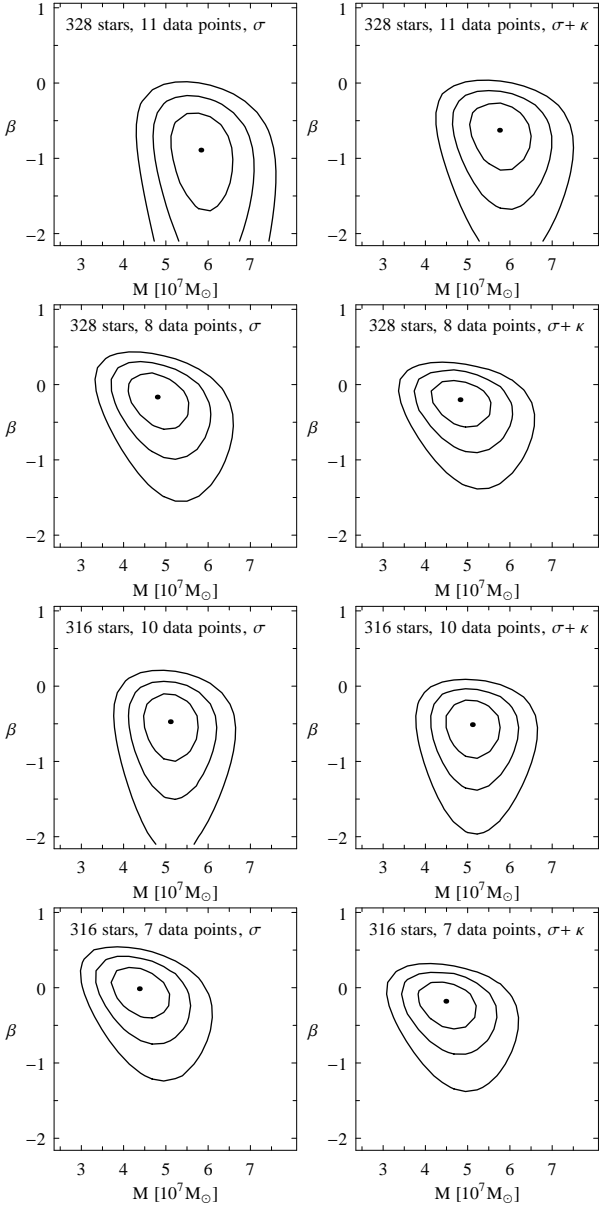}
\end{center}
\caption{The $1\sigma$, $2\sigma$ and $3\sigma$ confidence regions in the $M-\beta$ parameter
plane obtained from fitting the dispersion (left panels) and both dispersion and kurtosis (right
panels). The text in each panel specifies the sample for which the fit was performed. The best-fitting
parameters (marked with dots) are listed in Table~\ref{fitted} with $1\sigma$ error bars.}
\label{contours}
\end{figure}

Figure~\ref{moments} shows in the upper left panel the kinematic sample of 328 stars from M08.
The diagram plots the heliocentric velocities of Leo I stars as a function of distance from the
centre of the galaxy which we also adopt from M08 (see Table~\ref{adopted}). The selection
of these 328 stars out of the entire sample of 387 stars listed in table~5 of M08 was done by
rejecting obvious outliers including pronounced contribution from the Milky Way stars (see below).
From these data we calculated the velocity dispersion profile $\sigma(R)$ shown in the middle
left panel of Fig.~\ref{moments} in 11 radial bins of $9 \times 30 + 2 \times 29$ stars using
a standard unbiased estimator of dispersion (see e.g. {\L}okas, Mamon \& Prada 2005).
The data points were assigned sampling errors of size $\sigma/\sqrt{2(n-1)}$
where $n$ is the number of stars per bin. The lower left panel plots the kurtosis-like
variable $k = (\log \kappa)^{1/10}$ which has a Gaussian sampling distribution contrary to
the kurtosis $\kappa$. The values of $k$ were obtained with the correction of the standard estimator
of kurtosis $K$ by the bias due to the low number of stars per bin so that $\kappa=3 K/2.68$.
The data points were
assigned sampling errors of 2 percent (see {\L}okas \& Mamon 2003; {\L}okas et al. 2005).

For the modelling we made the simplest possible assumption that mass follows light or equivalently
that $M/L$ is constant with radius. The light distribution in terms of the
S\'ersic (1968) profile (for the formulae see {\L}okas et al. 2005) with $R_{\rm S}= 5.0$ arcmin
and $m=0.6$ was adopted from M08. The total apparent magnitude of Leo I was taken from
Irwin \& Hatzidimitriou (1995) and translated
into the absolute value using the distance of Leo I $D=254$ kpc as estimated by Bellazzini et al. (2004)
from the tip of the red giant branch (which agrees well with the distance found by Held
et al. 2001 from the RR Lyrae stars). The values of all the adopted parameters with errors are listed
in Table~\ref{adopted}. The error in luminosity includes the error in the measured apparent magnitude
as well as the distance.

We modelled the velocity moments using the solutions of the Jeans equations as described in
{\L}okas (2002) and {\L}okas
et al. (2005) adjusting two free parameters, the total mass and the anisotropy parameter $\beta$
which was assumed to be constant with radius. The best-fitting solutions in the case when only
the dispersion profile is considered are plotted as dashed lines in Fig.~\ref{moments} and the
corresponding confidence regions in the $M-\beta$ parameter plane following from the sampling
errors are illustrated in the left column of Fig.~\ref{contours}. The solid lines in Fig.~\ref{moments}
show the resulting best-fitting profiles in the case when both dispersion and kurtosis profiles are
fitted simultaneously. The corresponding confidence contours are plotted in the right column
of Fig.~\ref{contours}. For all cases the best-fitting parameters with $1\sigma$ errors are listed in
Table~\ref{fitted}.

\begin{table*}
\begin{center}
\caption{Fitted parameters of Leo I. }
\begin{tabular}{lccccc}
sample & fitted & $\beta$     & $M[10^7 $M$_\odot]$   & $M/L_V[$M$_\odot/$L$_\odot]$ & $\chi^2/N$ \\
\hline
328 stars, 11 data points & $\sigma$        & $-0.9^{+0.5}_{-0.8}$ & $5.8 \pm 0.7$ & $10.6 \pm 5.5$ & $21.0/9$  \\
			  & $\sigma+\kappa$ & $-0.6^{+0.3}_{-0.6}$ & $5.8 \pm 0.7$ & $10.5 \pm 5.5$ & $32.0/20$ \\ \\
328 stars, 8  data points & $\sigma$        & $-0.2^{+0.4}_{-0.4}$ & $4.8 \pm 0.7$ & $8.7 \pm 4.8$ & $3.0/6$   \\
			  & $\sigma+\kappa$ & $-0.2^{+0.3}_{-0.4}$ & $4.8 \pm 0.7$ & $8.8 \pm 4.8$ & $10.5/14$ \\ \\
316 stars, 10 data points & $\sigma$        & $-0.5^{+0.4}_{-0.5}$ & $5.1 \pm 0.6$ & $9.3 \pm 4.8$ & $13.6/8$  \\
			  & $\sigma+\kappa$ & $-0.5^{+0.3}_{-0.5}$ & $5.1 \pm 0.6$ & $9.3 \pm 4.8$ & $16.6/18$ \\ \\
316 stars, 7  data points & $\sigma$        & $\ \; \; 0.0^{+0.3}_{-0.4}$ & $4.4 \pm 0.7$ & $7.9 \pm 4.5$ & $1.5/5$   \\
			  & $\sigma+\kappa$ & $-0.2^{+0.3}_{-0.4}$ & $4.5 \pm 0.7$ & $8.2 \pm 4.5$ & $5.1/12$ \\
\hline
\label{fitted}
\end{tabular}
\end{center}
\end{table*}

The contours shown in the upper left panel of Fig.~\ref{contours} correspond to the result obtained
by M08, namely that if we force the assumption that mass follows light then the inferred anisotropy
will be rather tangential (with isotropy excluded at $3\sigma$ confidence).
In addition, the overall fit is quite bad (with $\chi^2/N=21.0/9$, see
Table~\ref{fitted}). This is obviously caused by the larger values of the outer 3 dispersion data
points. Interestingly, when the kurtosis is added to the analysis, the best-fitting anisotropy is
less tangential (see the upper right panel of Fig.~\ref{contours}) but the quality of the fit
is still bad ($\chi^2/N=32.0/20$). The secondary increase of the velocity dispersion profile at
larger projected radii $R$ may be interpreted as a signature of an extended dark matter halo but also
as due to the contamination by stars from the tidal tails if they are aligned with the observer's
line of sight.

As discussed in detail by Klimentowski et al. (2007), these stars will contribute
significantly to artificially inflate the velocity dispersion mainly in the outer radial bins.
If this is the case, it is advisable (as suggested by M08 themselves) to use only the inner part
of the velocity dispersion profile. We have therefore repeated the analysis rejecting the outer three
bins in the velocity dispersion and kurtosis profiles. The best-fitting solutions of the Jeans
equations for such truncated moments are plotted with thicker lines in Fig.~\ref{moments} ending at the
last point included in the fit. Again the dashed and solid lines correspond to the fits done for the
dispersion alone and for both moments. Note that the inferred anisotropy is now (second row of panels
in Fig.~\ref{contours}) consistent with isotropy at $1\sigma$ level, and the kurtosis only helps to
constrain the anisotropy more strongly.

\begin{figure}
\begin{center}
    \leavevmode
    \epsfxsize=8.4cm
    \epsfbox[105 55 475 410]{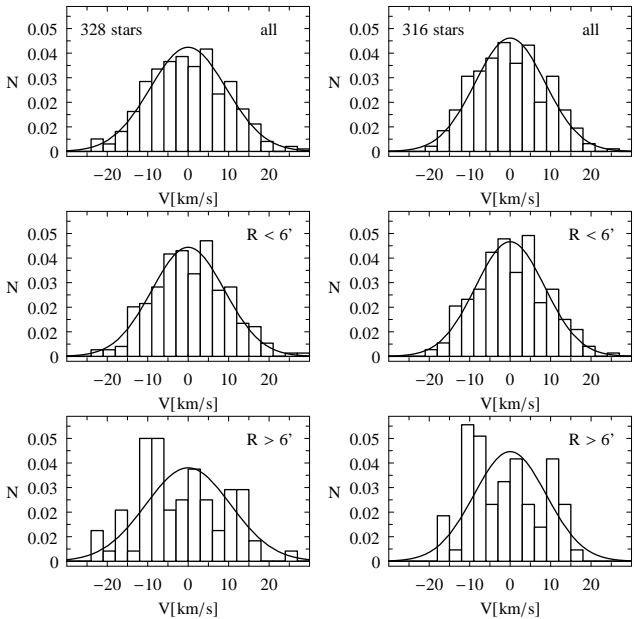}
\end{center}
\caption{The velocity distribution of Leo I stars. The left (right) column panels show the distribution for the
sample of 328 (316) stars. The upper panels are for the whole samples, the lower ones for the samples divided into
bins with $R<6$ arcmin and $R>6$ arcmin. The solid lines are Gaussian distributions with dispersions estimated
from velocities of stars in the corresponding bin.}
\label{histogramsleo}
\end{figure}

The need to restrict the analysis to the inner samples is
further supported by the analysis of the actual velocity distribution in Leo I. The left column
of Figure~\ref{histogramsleo} shows this distribution (normalized to unity) for all 328 stars (upper panel)
and separately inside and outside the projected radius of $R=6$ arcmin (middle and lower panel
respectively). The solid lines show the
Gaussian distributions with dispersions estimated from the data in a given bin. Although departures
from Gaussianity are expected for bound systems such as Leo I, at $R>6$ arcmin the distribution
is highly irregular making the estimates of velocity moments very uncertain.

Klimentowski et al. (2007) demonstrated that the contamination from the tidal tails can be at least
partially removed from the kinematic data sets by adopting the interloper removal method of
den Hartog \& Katgert (1996) originally devised for galaxy clusters. The method turned out to work
very effectively on mock kinematic data sets generated from a simulated dwarf galaxy being tidally
stripped by the Milky Way potential removing most of
unbound stars from the tidal tails present in the data due to projection effects. Applying this
method to the present sample for Leo I we remove 12 stars marked in the upper right panel of
Fig.~\ref{moments} as open circles. The corresponding velocity moments calculated for this reduced
sample of 316 stars are also shown in the right column of the Figure. We repeated the fitting of
the moments as for the entire sample and the results are shown in Figs.~\ref{moments} and \ref{contours}
as well as Table~\ref{fitted}
in an analogous way. We can see that the dispersion in the outer bins is now significantly reduced and
even when all 10 data points are fitted the quality of the fit is acceptable (only the outermost
dispersion point with $R>10$ arcmin is really discrepant). Note also that the
best-fitting masses are now somewhat lower than for the sample of 328 stars.

From the appearance of the velocity distribution for 316 stars (see the right column panels of
Fig.~\ref{histogramsleo}) it is again advisable to restrict the fit to the inner
7 data points of each velocity moment. In this case the quality is further improved and the results are fully consistent
with isotropy of stellar orbits and the hypothesis of mass following light. As the final results of our
analysis we suggest adopting those obtained for the most reliable sample of 7 inner data points for each moment
calculated from the set of 316 stars. From fitting both velocity moments we obtain in this case with
$\chi^2/N=5.1/12$ the anisotropy $\beta=-0.2^{+0.3}_{-0.4}$ and the total mass
$M=(4.5 \pm 0.7) \times 10^7$ M$_\odot$. The quoted errors are the
$1\sigma$ errors following from the sampling errors of velocity moments. Combining this mass with
the luminosity from Table~\ref{adopted} we get the mass-to-light ratio $M/L_V=8.2 \pm 4.5$ M$_\odot/$L$_\odot$
where the error includes the error in mass, the measurement of the apparent magnitude and the distance
(the values of $M/L_V$ for other fitted cases are listed in Table~\ref{fitted}).
Since the stellar mass-to-light ratio of the relatively young stellar population of Leo I
is estimated to be below 1 M$_\odot/$L$_\odot$ this value points to the presence of a significant
amount of dark matter. We conclude however that the kinematic data for Leo I can be explained without
an extended dark matter halo as the assumption of mass following light works quite well.

\begin{figure}
\begin{center}
    \leavevmode
    \epsfxsize=8.5cm
    \epsfbox[100 25 275 190]{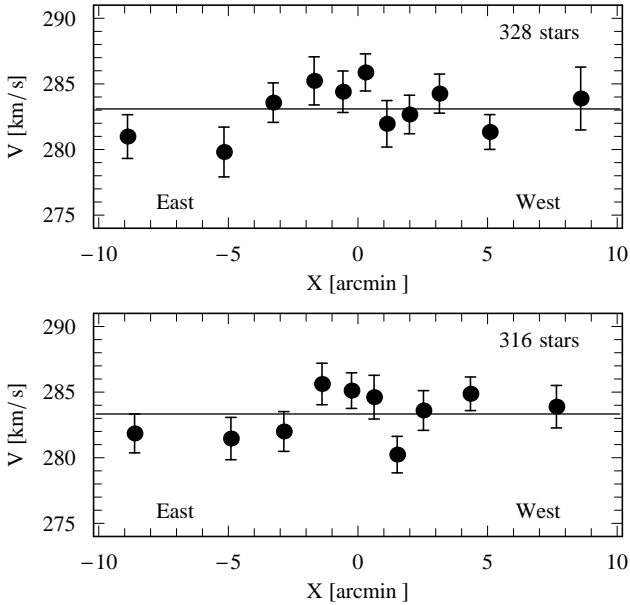}
\end{center}
\caption{Rotation curves of Leo I stars along the major axis of the photometric image. The upper panel shows
the original sample of 328 stars identified as members by M08, while the lower one presents the sample of 316
stars obtained after interloper rejection. The binning is $9 \times 30 + 2 \times 29$
for the sample of 328 stars and $4 \times 31 + 6 \times 32$ for the sample of 316 stars
(from negative to positive $X$). The horizontal line in each panel marks the mean velocity of each sample.}
\label{rotation}
\end{figure}

In order to verify our hypothesis that the kinematic data set for Leo I is indeed
contaminated by stars from the tidal tails we propose to consider the rotation curve obtained from the
same data. The curves are obtained by binning the velocities in a similar way as before but along the
major axis of the dwarf (assumed to lie at PA$=79^\circ$, as determined by Irwin \& Hatzidimitriou 1995).
The results are shown in Fig.~\ref{rotation} again for the total sample of 328 stars and the cleaned
sample with 316 stars. As we can see, the inner parts of the diagrams ($|X|<3$ arcmin) are consistent
with weak rotation such that the western part is approaching and the eastern receding. The direction of
the rotation is reversed however when we go farther out from the centre of the dwarf. Interestingly, this
result agrees well with what was reported in S07 (see their fig. 17) despite the fact that their data cover only a
fraction of the dwarf area on the sky (two differently oriented
rectangles on the eastern and western side of the dwarf, see their fig. 16). The rotation curve shown in
Fig.~\ref{rotation} also agrees with that of the eastern part shown in fig. 3 of K07  (in the western
part their data show no rotation).

\section{Comparison with $N$-body simulations}

\subsection{Rotation and tidal tails}

In order to verify whether such a reversed rotation can be due to tidal interactions we have resorted to an
$N$-body simulation. We used the last output of the simulation described in Klimentowski et al. (2007).
The simulation followed the evolution of a two-component (stars and dark matter) dwarf orbiting in the
Milky Way potential. Over 10 Gyr of evolution the dwarf completes five eccentric orbits losing $\sim 99$ percent of its
initial mass. The tidal interactions lead to the formation of pronounced tidal tails which are present for most
of the time. In the final output, the dwarf is at the apocentre, its shape is spheroidal, and the tidal tails are
oriented approximately towards the centre of the Milky Way (Klimentowski et al. 2008).

The orbital apocentre of the simulated dwarf
is 110 kpc which corresponds to a distance more than twice smaller compared to the current distance of Leo I.
Note however that a typical
cosmological orbit for satellites with apocentre to pericentre ratio $r_{\rm apo}/r_{\rm peri} \approx 5$ and
$r_{\rm apo}$ comparable to the current distance of Leo I
should still allow the transformation from a disk to a spheroid to be completed after about 10 Gyr (Mayer et al.
2001). The current orbit of Leo I might also be the
result of scattering from an orbit with much higher binding energy where
the transformation might have been much more efficient
with the original apocentre much smaller (Sales et al. 2007; M08).

\begin{figure}
\begin{center}
    \leavevmode
    \epsfxsize=8.3cm
    \epsfbox[0 20 370 370]{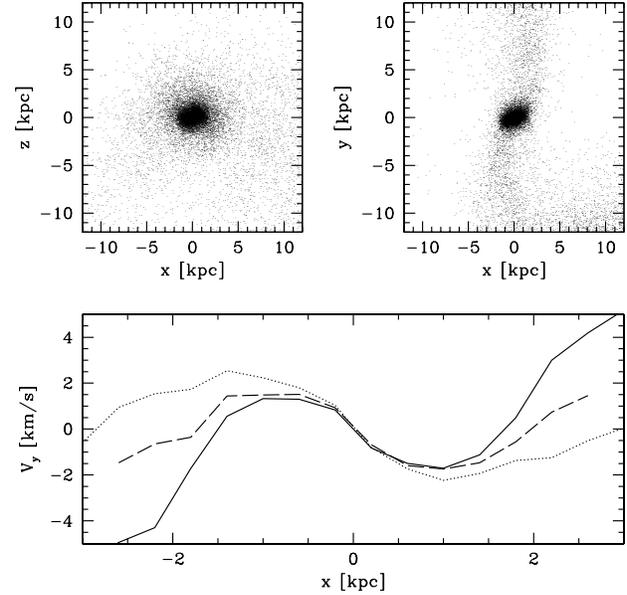}
\end{center}
\caption{Upper left: the simulated dwarf as seen on the sky. Upper right: the view of the dwarf from above. The
observer sees the galaxy along the $y$ axis ($x=0$) from below (more negative $y$).
Lower panel: the rotation curve measured by the observer from stars seen within $|x|<3$ kpc, $|z|<2.5$ kpc
and with velocities within $\pm 25$ km s$^{-1}$ with
respect to the dwarf's mean. The solid line shows the curve obtained with all stars, the dashed one from
the combined 100 samples of 200 stars after application of the interloper removal scheme. The dotted line
shows the rotation curve not affected by tidal tails, obtained by selecting stars within $|y|<2.5$ kpc.}
\label{sim}
\end{figure}

Given that dwarfs spend most of their orbital time near apocentre, the state of the simulated dwarf should be
{\it qualitatively\/} similar to that of Leo I. An observer situated near the Milky Way will view the dwarf
nearly along its tidal tails (Klimentowski et al. 2008). The configuration is
illustrated in Fig.~\ref{sim} with the upper left panel showing the dwarf as it appears to the observer on the
sky and the upper right panel showing the view from above the dwarf so that the line of sight is along the axis
$x=0$ and the observer is looking from the direction of negative $y$.
From such a configuration we choose stars with $|x|<3$ kpc and $|z|<2.5$ kpc and project their velocities along
the $y$ axis to produce line-of-sight velocities available for observation.
We also introduce a cut-off in these velocities at $\pm 25$ km s$^{-1}$ with respect
to the mean velocity of the dwarf, which corresponds to $\pm 4\sigma$ range in velocities where $\sigma \approx 6$
km s$^{-1}$ is the line-of-sight velocity dispersion of the stars in the centre of the dwarf. The solid line in the
lower panel of Fig.~\ref{sim} shows the rotation curve obtained from such
a data set by binning the data in a way analogous to the one applied to the data for Leo I. Note that our
simulated dwarf galaxy initially had a stellar disk which during the tidal evolution was transformed into a bar and
then to a spheroid. In the final stage some residual rotation is still present as verified in
Fig.~\ref{sim}. The equatorial plane of the dwarf (perpendicular to its total angular momentum vector)
is inclined by about 60 degrees to the orbital plane so we do not actually see the maximum rotation.

Fig.~\ref{sim} clearly demonstrates that a similar effect of reversed rotation as the one seen in Leo I can
be produced by the presence of strong tidal tails. The intrinsic rotation well visible in the inner part
of the simulated dwarf is reduced as we go towards larger $|x|$ and gets reversed at about $\pm 1.5$ kpc
which is well inside the dwarf. Note that the radius of the dwarf where the density profile starts to flatten
due to tidal tails is about 2.5 kpc (Klimentowski et al. 2007) which corresponds to about 10 arcmin for
Leo I. This behaviour is caused by the kinematics of stars in the tails: the stars typically move away from
the dwarf with velocities proportional to their distance (see fig. 21 of Klimentowski et al. 2007). In the
configuration presented in Fig.~\ref{sim} this motion is in the opposite direction with respect to the
intrinsic rotation of the dwarf on both sides of the galaxy.
In the inner parts the stars tracing the intrinsic kinematics of the dwarf
dominate and the rotation is well visible. When moving away from the centre there is a point where
the stars from the tidal tails start to prevail and the dominant motion changes direction.

In choosing the stars to calculate the rotation curve we made only a simple constant cut-off in velocity
with respect to the mean velocity of the dwarf. Although the sample of 328 stars in M08 was obtained in
a similar way, in Fig.~\ref{rotation} we demonstrated that the behaviour of the rotation curve is preserved
also for the sample of 316 stars cleaned with our method of interloper rejection. In order to check whether
this is the case also for the simulated data we randomly selected 100 samples of 200 stars each from the
total sample of stars used before and cleaned them of interlopers. The dashed line in Fig.~\ref{sim} shows
the average rotation curve calculated from these cleaned samples. As expected, the effect of the tidal tails
is now less pronounced, but the reversed rotation is still present. We verified using the full 3D information
from the simulation that the result shown with the dashed line would be almost identical if we actually removed
all unbound stars. The reason for this is that some of the stars in the tails are still bound to the dwarf
while moving away from it. On the other hand, if the tidal tails are cut off by considering only the stars
within $|y|<2.5$ kpc, the reversed rotation disappears, as shown by the dotted line in Fig.~\ref{sim}. This
proves that indeed the tidal tails are responsible for the reversal of rotation.

Although in Fig.~\ref{sim} the rotation is reversed at around 1.5 kpc (or 1.8 kpc for the cleaned sample)
corresponding to 0.6 of the dwarf radius, in Leo I it seems to occur at 3 arcmin,
i.e. at a much smaller fraction of radius equal to 0.3. Such diffferences are expected in light of
the fact that the simulated dwarf was never intended to be a precise model of Leo I. It is simply employed to
propose a plausible model for the kinematics of Leo I. The radius of rotation reversal could
be easily changed by varying simulation parameters such as the initial concentration
of the halo or the initial disk scale length of the stars which will
affect the effective tidal radius (Mayer et al. 2002).

If the overall qualitative picture presented here is
correct then the eastern side of Leo I must be affected by the trailing tidal tail
while the western side by the leading tail. Note that the rotation curves for 328 stars and 316 stars
(upper and lower panels of Fig.~\ref{rotation}) are quite similar. This further confirms our suspicion from the
previous section that the sample of 316 stars is still to some extent contaminated by tidal tail stars. This
contamination, however subtle, should still be taken into account; it supports our suggestion to include only the
inner data points of the dispersion and kurtosis in the kinematic modelling of Leo I.

\subsection{The symmetry of the velocity distribution}

\begin{figure}
\begin{center}
    \leavevmode
    \epsfxsize=8.4cm
    \epsfbox[105 55 475 415]{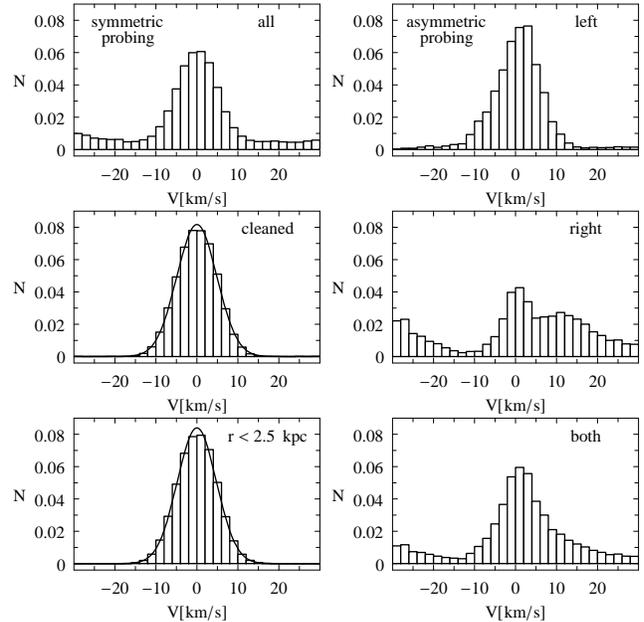}
\end{center}
\caption{Upper left panel: the distribution
of velocities of $1.2 \times 10^5$ stars along the $y$ axis in the same configuration as shown in Fig.~\ref{sim}.
Middle left panel: a similar distribution obtained from a sum of 100 samples
of 200 stars selected randomly and cleaned of interlopers (in the end we
have $1.6 \times 10^4$ stars). Lower left panel: velocity distribution of $7.7 \times 10^4$ stars
from the inside of the dwarf ($r<2.5$ kpc). In the two lower panels
the lines show the Gaussian distributions with dispersions measured from the data.
Upper (middle) right panel: the distribution of velocities of about 4000 stars from the left
horizontal (right vertical) window in Fig.~\ref{plotpos}. Lower right panel: the distributions from the left
and right windows combined. All histograms were normalized to unity.}
\label{histograms}
\end{figure}

Another issue discussed by S07 and M08 is the question of the symmetry of the velocity distribution of the
stellar sample of Leo I. While S07 found their distribution to be highly asymmetric, the one reported by
M08 was quite symmetric and Gaussian-like. In the upper left panel of Fig.~\ref{histograms} we show the distribution
of velocities of stars along the $y$ axis as measured by an observer situated in the same
way with respect to the dwarf as before, now with the $\pm 30$ km s$^{-1}$ cut-off in velocity, corresponding
to $5 \sigma$ range, larger this time to explore the tails of the velocity distribution.
Although the distribution for the dwarf stars is embedded in a uniform background from the tails, it appears quite
symmetric. In the middle left panel we plot a similar distribution but now obtained from a sum of 100 samples
of 200 stars each selected randomly from the previous one and cleaned of interlopers.
On top of the distribution we plotted a Gaussian with dispersion of $\sigma=
5$ km s$^{-1}$ calculated from the sample. The lower left panel shows the distribution of line-of-sight velocities
of bound stars from the inside of the dwarf, i.e. with radii $r<2.5$ kpc. Again, a Gaussian
with $\sigma=4.9$ km s$^{-1}$ calculated from the sample is plotted on top.

\begin{figure}
\begin{center}
    \leavevmode
    \epsfxsize=7.4cm
    \epsfbox[90 30 290 190]{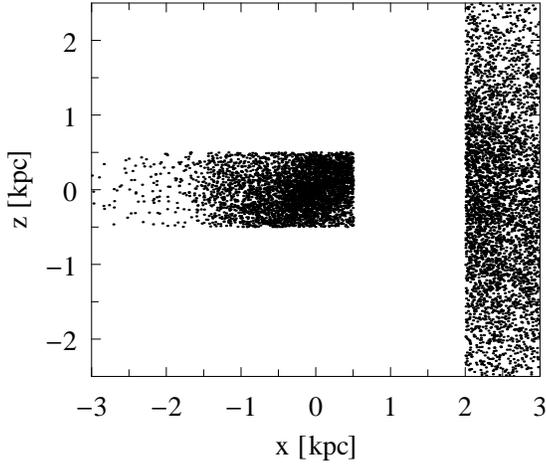}
\end{center}
\caption{The selection of stars for the asymmetric probing. The configuration and line of sight is the same
as in Fig.~\ref{sim}. The left horizontal window has $|z|<0.5$ kpc and $-3$ kpc $< x < 0.5$ kpc, the right
vertical one has $|z|<2.5$ kpc and $2$ kpc $< x < 3$ kpc. Both windows have about 4000 stars.}
\label{plotpos}
\end{figure}

In all three cases, the distribution
is highly symmetric which is understandable because in the present configuration the tidal tails
contribute to both negative and positive velocities similarly. In addition, the two lower panels show remarkable
similarity which means that our interloper removal scheme works adequately in removing obvious outliers
(i.e. the uniform distribution of the stars from the tails). However, it should be kept in mind that the
method does not remove all interlopers. Indeed, in a configuration similar to the one used here (observation along
the tidal tails) and with the same initial cut-off in velocity, the scheme removes on average 80 percent of unbound
stars (Klimentowski et al. 2007). The remaining contamination, although not apparent in overall distributions
like the ones shown in the left column of Fig.~\ref{histograms}, is still present and responsible for
the reversed rotation as demonstrated by Fig.~\ref{sim}.

We have shown that in the proposed configuration the underlying velocity distribution should be symmetric.
Whether this is actually seen in the data will depend however on the uniformity of probing. While the stars selected
for spectroscopic measurements by M08 were uniformly distributed across Leo I, the sample of S07 probed two very
different regions. In order to mimic their observations we selected the stars in a similar way, shown in
Fig.~\ref{plotpos}. The cut-off in velocity $V_y$ was $\pm 30$ km s$^{-1}$ as before. The velocity distribution of stars
from the left horizontal window with $|z|<0.5$ kpc and $-3$ kpc $< x < 0.5$ kpc is shown in the upper right
panel of Fig.~\ref{histograms}. It is close to symmetric, because this region is dominated by the stars from the
inside of the dwarf, with a small excess of stars with negative velocities contributed by the trailing tidal tail.
The velocity distribution from the right vertical window with $|z|<2.5$ kpc and $2$ kpc $< x < 3$ kpc
is plotted in the middle right panel of Fig.~\ref{histograms}. In this case the distribution is dominated by positive
velocities of the stars in the leading tidal tail. Since in the data of S07 both windows had a comparable
number of stars we reduced the sample of the left window by a factor of 10 to have a similar number of stars in both
windows. The combined velocity distribution from both windows is shown in the lower panel of Fig.~\ref{histograms}.
It is significantly asymmetric with an excess of stars with positive velocities from the leading tidal tail.
This shows clearly that the asymmetric distribution of stars found by S07 may be due to the asymmetric probing
of a symmetric parent distribution and not necessarily to the bias in their velocity measurements as suggested
by M08.

\subsection{Tests of the method of velocity moments}

\begin{figure}
\begin{center}
    \leavevmode
    \epsfxsize=8.2cm
    \epsfbox[110 50 470 410]{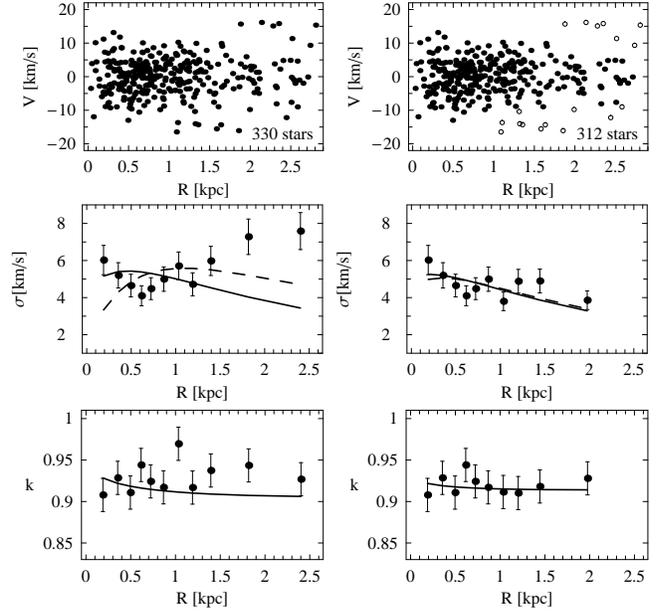}
\end{center}
\caption{Upper panels: the kinematic data sets for sample 10 generated from the simulated dwarf. The left panel shows
the original sample of 330 stars selected by the cut-off in velocity of $\pm 17$ km s$^{-1}$ with respect to
the dwarf's mean velocity, while the right one presents the sample of 312
stars obtained by rejection of interlopers (open circles). The middle (lower) panels show the velocity
dispersion (kurtosis) profiles obtained from the corresponding samples with 30 stars per bin. The dashed lines
show the best-fitting dispersion profiles when only the dispersion is fitted, while the solid lines plot the
best-fitting profiles of the moments when the dispersion and kurtosis are fitted simultaneously.}
\label{sample10}
\end{figure}

We used the simulation data in the same configuration to test the reliability of the method used to model Leo I,
based on fitting the velocity dispersion and kurtosis profiles. The method has been tested extensively
in the context of modelling the kinematic samples of galaxy clusters and shown to reproduce well the properties
of simulated cluster-size dark matter haloes (Sanchis, {\L}okas \& Mamon 2004; {\L}okas et al. 2006, 2007).
Here we proceed in a similar way and generate 10 kinematic samples from the stellar component of our simulated
dwarf of 330 stars each. The stars were selected randomly from the region corresponding to a projected radius of
$R < 3$ kpc and with an initial cut-off in velocity of $\pm 17$ km s$^{-1}$ with respect to the dwarf's mean velocity,
which corresponds to $3\sigma$ range, exactly as in the original data set of M08.
An example of such data set is shown in the upper left panel of
Fig.~\ref{sample10} for sample number 10 in a way analogous to the way we presented the data for Leo I in
Fig.~\ref{moments}.

\begin{figure}
\begin{center}
    \leavevmode
    \epsfxsize=8.2cm
    \epsfbox[110 50 470 410]{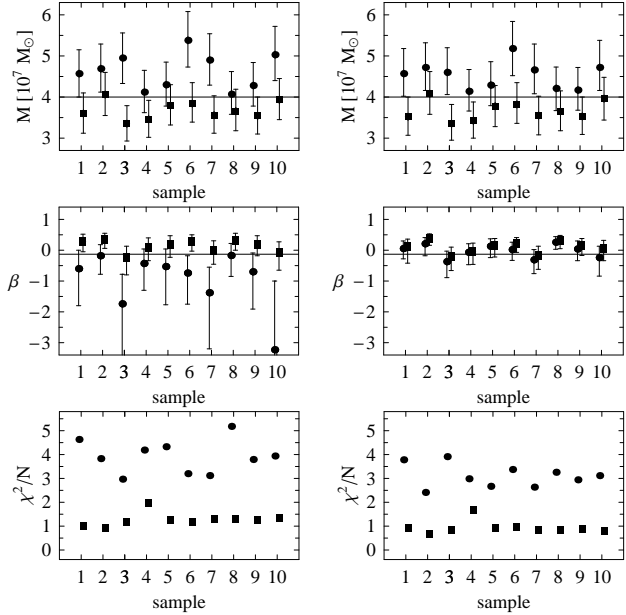}
\end{center}
\caption{Results of fitting the velocity moments for
10 mock data samples generated from the simulated dwarf. The left panels show the
best-fitting mass (upper panel) and anisotropy (middle panel) with $1\sigma$ errors and the goodness of
fit measure (lower panel) in the case when only the dispersion profile is fitted. The right panels show the
corresponding results in the case when both moments are fitted. The circles correspond to the results obtained
for the entire samples, while the squares to those for samples cleaned of interlopers. Horizontal solid lines indicate
the true values of the parameters measured from the 3D information.}
\label{samples}
\end{figure}

The data were binned with 30 stars per bin to obtain the profiles of the velocity moments also shown
in Fig.~\ref{sample10}. We then fitted the moments with the solutions of the Jeans equations adopting the
assumptions that mass follows light and anisotropy $\beta=$const. The distribution of light was obtained by
fitting the S\'ersic profile to the projected distribution of the stars which gave $R_S = 0.54$ kpc and $m=1$.
The errors on the estimated parameters, the total mass and anisotropy,
were read from probability contours analogous to the ones for Leo I
shown in Fig.~\ref{contours}. The best-fitting parameters together with $1\sigma$ errors are presented in
Fig.~\ref{samples} for all samples, including sample 10. The horizontal solid lines in the four panels
in the upper two rows of the Figure indicate the true values of the parameters measured from the full 3D information
on the simulated dwarf: $M = 4.0 \times 10^7$ M$_\odot$ and $\beta=-0.13$ (see Klimentowski et al. 2007).

The left panels of Fig.~\ref{samples} show results for the case when only the velocity dispersion profile
is fitted. When the entire samples of 330 stars are considered (circles), the quality of the fit is generally
poor. More specifically, compared to the true properties of the dwarf measured from the 3D simulation data
(horizontal solid lines), the mass is overestimated and the anisotropy underestimated. In particular, in three
cases out of ten, the best-fitting value of $\beta$ is $\beta < -1$, including the most discrepant case of
sample 10 which has $\beta=-3.2$. A similar low anisotropy
was obtained for Leo I when the data were treated in the same way.
When the data are cleaned of interlopers (we then use 300 stars in 10 bins) and again only the velocity
dispersion profile is fitted (squares), the quality of the fit improves dramatically. The masses are now only
slightly underestimated and anisotropies slightly overestimated
which is due to the specific properties of the velocity distribution in the dwarf in this configuration (see
Klimentowski et al. 2007).

\begin{figure}
\begin{center}
    \leavevmode
    \epsfxsize=8.4cm
    \epsfbox[105 55 470 410]{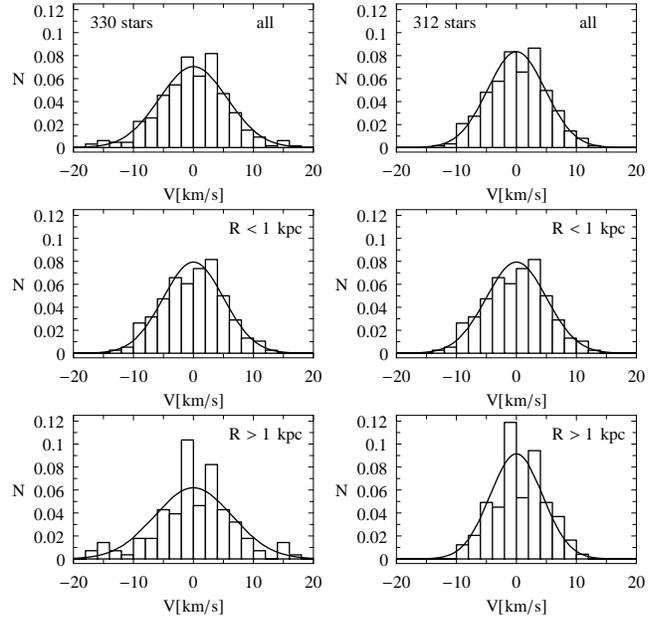}
\end{center}
\caption{The velocity distribution of stars in sample 10. The left (right) column panels show the distribution for the
sample of 330 (312) stars. The upper panels are for the whole samples, the lower ones for the samples divided into
bins with $R<1$ kpc and $R>1$ kpc. The solid lines are Gaussian distributions with dispersions estimated
from velocities of stars in the corresponding bin.}
\label{histsample10}
\end{figure}

The right panels of Fig.~\ref{samples} show the corresponding results in the case when both velocity dispersion and
kurtosis are fitted. As expected, the mass estimates are similar to those in the left panel. This is because it is
the dispersion profile which is mainly sensitive to the mass. However, the situation is completely different for the
anisotropy. Although for the entire samples (with interlopers) the quality of the fits is still poor, the best-fit
anisotropy values are in close agreement with the true 3D values. Interestingly, analysis of both types of samples
result in very similar values of anisotropy suggesting that contamination does not affect the determination of
$\beta$ when both dispersion and kurtosis are fitted. This is due to the fact that, contrary to common belief,
the kurtosis is not much more affected by contamination than the dispersion. The reason for the
similar effect of interlopers on kurtosis may be that it is constructed via dividing the fourth velocity
moment by the fourth power of dispersion so the influence of contamination may partially cancel out.
The contamination increases the measured values of kurtosis (see the lower panels of
Fig.~\ref{sample10}), but the dispersion and the kurtosis depend on the velocity anisotropy in a different way,
so that more strongly increasing dispersion corresponds to more tangential orbits
while more strongly increasing kurtosis corresponds to more radial orbits (see fig. 4 in {\L}okas et al. 2005).
The kurtosis values inflated by interlopers thus lead to the situation when more radial orbits
are preferred which balances the preference for tangential orbits due to the inflated dispersion profile.

In the case when both moments are fitted
for samples cleaned of interlopers the overall quality of the fits is good and in eight out of ten cases the true values
of the parameters are within the $1\sigma$ errors of the best-fitting parameters. We can thus be quite confident
that the error estimates for Leo I reflect the real uncertainty in the parameters. Note that the velocity distribution
of stars from the simulated dwarf appears more regular than for the real sample of Leo I (see Fig.~\ref{histsample10})
so there is no need to restrict the analysis to the inner part.

\section{Discussion}

We have demonstrated that after the application of the interloper removal scheme and the inclusion of the fourth
velocity moment the kinematic data for Leo I dSph galaxy can be reliably modelled using the solutions
of the Jeans equations. We find that the data are consistent with a simple model
in which mass follows light and stellar orbits are close to isotropic with no need for an extended dark matter halo.
The picture of Leo I emerging from our analysis is fully consistent with the tidal stirring scenario for the
formation of dSph galaxies (Mayer et al. 2001). The fact that Leo I has been significantly tidally stirred
and that fairly prominent tails are present is not in contradiction with the fact that it is still a gravitationally
bound object nearly in equilibrium and has a significant dark matter halo. This dark matter halo is not extended, but
rather truncated just outside the stellar component. The scenario predicts that extended dark haloes are removed
after one or two tidal shocks during pericentre passages if only the pericentre is small enough, i.e. below 50 kpc
(Mayer et al. 2001, 2002).
Also the intrinsic rotation present in Leo I is naturally explained within this scenario as a remnant of the initial disk.

The study of the rotation curve in Leo I shows that the
cleaned sample is still contaminated to some extent by tidal tails and therefore it is advisable to use only the
kinematic data from the inner part of the dwarf for dynamical modelling. The rotation curve constructed from the data,
both for the initial and cleaned samples, shows that the rotation is reversed when going from the inside to the outside
of the galaxy. We interpret this behaviour as due to the presence of contamination from the leading tidal tail
in the western part of Leo I (seen in the background from the point of view of the observer) and from the trailing
tail in the eastern part of the galaxy (seen in the foreground of the observer).

This kind of behaviour in the rotation curve may also reflect the presence of a counterrotating core.
Such cores are however typically
found in much brighter elliptical galaxies; an example is the galaxy NGC 770 studied in detail by
Geha, Guhathakurta \& van der Marel (2005). The rotation curve of the galaxy, shown in their fig. 3 looks very
similar to the rotation curve of Leo I in Fig.~\ref{rotation} of the present paper and the photometric
analysis shows that it is
generated by the presence of a small inner disk. For the inner rotation of the disk to be well visible it has to
be viewed close to edge-on. Then the inner contours of the surface brightness should be more disky than the outer
ones, i.e. the ellipticity should decrease with radius. It is indeed the case for NGC 770, as demonstrated by
fig. 7 of Geha et al. (2005), but not for Leo I: as shown in fig. 17 of M08 the ellipticity of Leo I increases
with radius, i.e. the inner contours of the surface density of the stars are more circular. This could only be
reconciled with counterrotation if there is a counterrotating bar in Leo I viewed along the long axis.
It is unclear how such a bar could form in the standard tidal stirring scenario for the formation of
dSphs (Mayer et al. 2001). Although in the simulation employed here
the dwarf has a bar for most of the time (it is destroyed only at the last pericentre, see Klimentowski et al.
2008), it rotates in the same direction as the rest of the stars. One possibility of creating a
counterrotating bar is through an interaction of Leo I with some other dwarf galaxy in the past
(e.g. Kravtsov, Gnedin \& Klypin 2004).

In light of the analysis presented here, Leo I looks very similar to the Fornax dwarf for which it is also found
that after the removal of interlopers the model with mass following light and stellar orbits close to
isotropic provides satisfactory description of the kinematics (Klimentowski et al. 2007; {\L}okas,
Klimentowski \& Wojtak 2007). Both dwarfs also have a rather low mass-to-light ratio compared to systems like Draco.
The important difference is the source of contamination in the kinematic samples: while in Fornax
the majority of contamination probably comes from Milky Way stars, in Leo I it is due to tidal tails.
In the direction of Leo I the contamination from the Milky Way is negligible. According to the Besancon model
of the Milky Way (Robin et al. 2003),
the stars from our Galaxy are expected to have heliocentric velocities below 100 km s$^{-1}$. This
corresponds roughly to a 200 km s$^{-1}$ difference with respect to the mean velocity of Leo I stars.

Our estimates of the mass and mass-to-light ratio agree within errors with those of M08 and S07 but are by
a factor of a few lower than those of K07. Once the difference in the (much lower) assumed
luminosity of K07 is taken into account, the discrepancy is alleviated (K07 assumed $L_V= 3.4 \times 10^6$ M$_\odot$
which should have been corrected for the much lower distance adopted by Irwin \& Hatzidimitriou 1995).
We fitted the dispersion profile from the lower right panel of fig. 8 in K07 with
our adopted parameters (see Table 1) and assumptions (mass follows light and $\beta$=const). We find that our
model fits their data well for a value of mass as low
as $(7.3 \pm 2.1) \times 10^7$ M$_\odot$ which corresponds to $M/L_V=(13.3 \pm 9.1)$ M$_\odot/$L$_\odot$.
This value differs by $\sim 60$ percent from our preferred value of $M/L_V=8.2$, but the two agree within
$1\sigma$ errors.

Comparing our velocity dispersion profile from M08 (even for the original contaminated sample)
to that of K07, we find that the latter is typically higher (note however that, contrary to the data sets
of M08 and S07, no secondary increase of the dispersion profile at larger radii is seen).
This should not be due to errors in velocity measurements which
are of the order of 5 km s$^{-1}$ in K07, compared to 2 km s$^{-1}$ in M08, because K07 used the maximum likelihood
estimator of dispersion which should have taken them into account. However, when estimating
their velocity dispersion profile they combined data sets from different instruments and applied a rather
conservative approach to interloper rejection (i.e. only $3\sigma$ outliers were rejected which is not
sufficient, see Klimentowski et al. 2007, Wojtak et al. 2007). Both these factors could significantly inflate
the velocity dispersion profile.

Using an $N$-body simulation we have shown that the line-of-sight velocity distribution of a dwarf near apocentre
should be symmetric and close to Gaussian in shape, as is indeed the case for the stellar sample of M08 for Leo I.
The reason why S07 found the distribution of their sample
to be asymmetric probably lies in their non-uniform coverage of the galaxy or the bias in velocity measurements.
Leaving aside the asymmetry, the main difference between the interpretation of the Leo I data proposed by S07 and
the one proposed by M08 and here lies in the type of tidal debris responsible for the velocity gradient in the
rotation curve. While
here we proposed that the gradient is due to recently formed tidal tails in the immediate vicinity of the dwarf,
S07 claimed that
it is due to large scale tidal streams formed earlier and approximately following the orbit.
When the dwarf is approaching the
apocentre, as Leo I likely does, the leading stream should slow down (because it approaches the apocentre earlier)
and the trailing stream should speed up with respect to the dwarf. This could result in an opposite behaviour of the
streams with respect to the tails near the dwarf (the latter are formed from matter moving away from the dwarf in all
parts of the orbit) and an opposite assignment of leading and trailing debris than the one proposed here. We can think
of four arguments why this interpretation of the data seems less likely than ours:
\begin{enumerate}
\item As pointed out by M08, any contamination from tidal debris
in Leo I probably comes from regions close to the dwarf because the usual features seen in the
colour-magnitude diagram of
Leo I are well visible (they would be blurred if the stars came from a wide variety of distances).
Based on photometric measurements, M08 estimate the
tidal extensions of Leo I to correspond to distances less than 40 kpc.
When studying the velocity distribution in the $N$-body simulation,
we introduced a velocity cut-off of $\pm 30$ km s$^{-1}$ which corresponds to distances of tidal tail stars
less than 20 kpc (see fig. 21 in Klimentowski et al. 2007), well within the allowed range.
On the other hand, significant contamination from the
large-scale tidal streams would require going to much larger distances.
\item For the large-scale tidal streams to contribute significantly, the observation would have to be
performed almost along
the orbit. For this to be possible from the inside of the Milky Way, the orbit would have to be very elongated, with
$r_{\rm apo}/r_{\rm peri}$ much larger than the typical values of the order of 5 found in cosmological simulations.
On the other hand, near apocentre the tidal tails in the vicinity of the dwarf are typically oriented
radially towards the Milky Way (see Klimentowski et al. 2008).
\item The density of the tidal debris must be very high in order to cause the inversion of the rotation curve in Leo I
well inside the dwarf. This density is highest in the tidal tails recently formed in the immediate vicinity of the dwarf.
\item Although the underlying velocity distribution should be symmetric, the inversion of the rotation velocity in
Leo I is better visible on the eastern side of the dwarf. This may be due to an observational bias related to the fact
that the selection of stars for spectroscopic measurements is made by introducing cuts e.g. in magnitude.
Since the stars in the trailing tail are closer to the observer and therefore appear brighter, they will be more
likely chosen than those from the leading tail. This further suggests
that the trailing tail is on the eastern side of Leo I as in our proposed configuration.
\end{enumerate}

\section*{Acknowledgements}

We are grateful to M. Mateo et al. for providing
the kinematic data for Leo I stars in electronic form. We also wish to thank S. Majewski and an anonymous
referee for comments which helped to improve the paper.
SK is funded by the Center for Cosmology and Astro-Particle Physics at The Ohio State University.
The numerical simulations were performed on the
zBox1 supercomputer at the University of Z\"urich.
We made use of the Besancon Galaxy model available at
http://bison.obs-besancon.fr/modele/.
This research was partially supported by the
Polish Ministry of Science and Higher Education
under grant N N203 0253 33.

\end{document}